\documentclass{aa}  

\usepackage{graphicx}
\usepackage{txfonts}
\usepackage{natbib}

\usepackage{longtable}

\begin{document}

   \title{Dynamical formation of detached trans-Neptunian objects close to the 2:5 and 1:3 mean 
motion resonances with Neptune}

   \author{P.I.O. Brasil
          \inst{1}
          ,
          R.S. Gomes\inst{2}
          ,
          \and
          J.S. Soares\inst{2}}

   \institute{Instituto Nacional de Pesquisas Espaciais (INPE), ETE/DMC,
              Av. dos Astronautas, 1758, S\~ao Jos\'e dos Campos, Brazil\\
              \email{pedro\_brasil87@hotmail.com}
         \and
             Observat\'orio Nacional (ON), GPA, Rua General Jos\'e Cristino, 77, Rio de Janeiro, Brazil\\
             \email{rodney@on.br}
	}

   \date{Received month/day, year; accepted month/day, year}

 
  \abstract
   {}
    { It is widely accepted that the past dynamical history of the solar system included a scattering of planetesimals from a primordial disk by the major planets. The primordial scattered population is likely the origin of the current scaterring disk and possibly the detached objects. In particular, an important argument has been presented for the case of $2004XR_{190}$ as having an origin in the primordial scattered disk through a mechanism including the 3:8 mean motion resonance (MMR) with Neptune (Gomes 2011). Here we aim at developing a similar study for the cases of the 1:3 and 2:5 resonances that are stronger than the 3:8 resonance.}
   {Through a semi-analytic approach of the Kozai resonance inside a MMR, we show phase diagrams ($e, \omega$) that suggest the possibility of a scattered particle, after being captured in a mean motion resonance with Neptune, to become a detached object. We ran several numerical integrations with thousands of particles perturbed by the four major planets, and there are cases with and without Neptune's residual migration. These were developed to check the semi-analytic approach and to better understand the dynamical mechanisms that produce the detached objects close to a MMR.}
   {The numerical simulations with and without a residual migration for Neptune stress the importance of a particular resonance mode, which we name the hibernating mode, on the formation of fossilized detached objects close to  MMRs. When considering Neptune's residual migration we are able to show the formation of detached orbits. These objects are fossilized and cannot be trapped in the MMRs again. We find a ratio of the number of fossilized objects with moderate  perihelion distance ($35<q<40au$) to the number of objects with high perihelion distance ($q>40au$) as 3.0/1 for objects close to the 2:5, and 1.7/1 for objects close to the 1:3 resonance. We estimate that the two fossilized population  have a total mass between $0.1$ and $0.3$ Pluto's mass. }
   {}


    \keywords{Kuiper Belt: general --
	      minor planets, asteroids: general
		}

\authorrunning {P.I.O. Brasil et al.}
 \titlerunning{Dynamical formation of detached TNOs close to the 2:5 and 1:3 MMRs with Neptune}

   \maketitle
%

\section{Introduction}

Since the mid-twentieth century, scientists have speculated about the existence of small bodies beyond
 the orbit of Neptune \citep{edgeworth1949, kuiper1951, kuiper1974}. These objects would form a disk and their 
orbits would have low eccentricities and inclinations. More than four decades later \citet{jewitt1993}
found 
the first object, 1992QB$_1$, (except for Pluto and Charon) belonging to the Kuiper belt. 
This object's orbital characteristics were consistent with those proposed by the first idealizers,
 with fairly low eccentricity and inclination. 
However, over the years, hundreds of new discoveries have revealed a much more complex scenario. One of the important unexplained
features was the unexpectedly large number of objects with high inclinations. 
The first dynamical formation models \citep{malhotra1993,malhotra1995} were very successful at explaining several features of
the Kuiper belt orbits, including the Plutinos and other resonant
orbits, but in the end they failed to explain
the high inclination of the Kuiper belt objects satisfactorily.
 Although up to now no theoretical model explain all the features of the trans-Neptunian 
region, the fact that there are many objects with high inclination orbits is something 
that differs a lot from what was initially expected. \citet{gomes2003a, gomes2003b, gomes2011} have introduced, through a disk scattering mechanism, a new
approach in producing high-inclination objects for the the hot population of the classical Kuiper belt and detached TNO's.

The detached objects are roughly defined as having  $ a>50 au$, $ q \gtrsim 36au$, and usually high inclination ($i>10^{\circ}$). These characteristics provide much more 
stable dynamics without close encounters with Neptune, as is the case of the scattering
disk objects whose perihelia, 
 {\it q}, of the orbits approach Neptune ($30<q<35au$) and have semimajor axis $a>50au$. The detached population currently has the 
lowest number of discovered members, owing to the 
difficulty  observing and tracking objects with large perihelia and  
high inclinations.  But, detecting, tracking, and understanding how
these objects could have formed may, nevertheless, reveal important 
dynamical processes that occurred in the primordial solar system \citep{gladman2002, 
gomes2005b, gomes2008, allen2006}.

There have been several hypotheses for the dynamical formation of detached objects belonging to the 
detached group,
 since they started to be discovered. The object $2000CR_{105}$ ($a=228.8au, q=44.13au, i=22.77^{\circ}$) is 
one of the first discoveries belonging to the detached objects group, called ``extended scattered
disk'' (ESD) at that time.
\citet{gladman2002} show that a process like long-term diffusive chaos is not good enough
to reproduce such orbit. Then they propose other alternatives, such as perturbations by 
distant rogue planets or by 
primordial stellar passages as possible candidates to explain $2000CR_{105}$ dynamical formation.
 \citet{gomes2006} show a mechanism for producing distant high-{\it q} detached objects, 
invoking the possibility that these bodies had interacted with a planetary-mass solar companion.
\citet{brasser2006, brasser2012a}, in turn, propose that 
the inner Oort cloud objects (actually, it is difficult 
to say when the detached population ends and the inner Oort cloud begins) would be formed by the perihelion-raising effect of a promordial cluster that was supposed to be the Sun's birth environment. The authors consider close stellar encounters
and the gas potential in the cluster as a scenario that produces such orbits with $a>200au$.

 \citet{gomes2003a, 
gomes2003b} and \citet{gomes2005b} present a scenario where detached 
objects are generated exclusively through gravitational interaction with the Jovian planets. 
The authors propose that objects scattered during planetary migration  can be captured 
in a mean motion resonance (MMR) with Neptune. 
Once in MMR, some particles could enter the Kozai resonance and present large variations in the 
eccentricities and inclinations. When the particle is in a low-eccentricity mode 
(and high inclination), the MMR critical angle may present a very large libration amplitude. 
Therefore, the object can escape both resonances while Neptune is still migrating,  
becoming a fossilized object with a high perihelion distance. Thus it is possible to produce 
detached  orbits belonging to the ESD by just considering the perturbation of the giant 
planets, at least  for $a \lesssim 100au$.  To fossilize the orbits it is only
necessary that Neptune experiences a residual migration. This mechanism could thus be valid in principle for both smooth migration 
\citep{hahn1999, hahn2005} as well as for ``direct emplacement'' migration models 
\citep{tsiganis2005, gomes2005a, nesvorny2011, nesvorny2012}. 

\citet{gomes2011} uses this mechanism to 
explain the dynamical formation of $2004XR_{190}$ which is a  fossilized object with high 
perihelion distance ($q\sim51.6au$) and 
 that is close to but not in the 3:8 MMR with Neptune \citep{allen2006}. In his
work the author highlights an interesting resonant mode in which the particle spends a long time with 
low eccentricity and high inclination (hibernating mode) and suggests that similar behavior 
should also occur for other MMRs. Although other models propose the formation of detached objects \citep {hahn2005}, we understand that the results in \citet{gomes2011} are compelling enough that we should reconsider the same mechanism applied to other important exterior MMRs with Neptune.

We thus 
intend to carry out  a comprehensive study in this paper of the dynamical 
formation of detached objects near the 2:5 and 1:3 MMRs with Neptune, which are the strongest ones beyond the 1:2 MMR, since they have the smaller combination of order and degree. We performed several numerical 
integrations 
of the four giant planets and two disks of scattered particles surrounding the resonances of 
interest. 
We considered scenarios with the actual solar system and another one where Neptune is  
a little before its current position, so that one can impose a
final phase of  residual migration on Neptune. Therefore we intend to increase our understanding of the dynamical 
formation mechanism of detached objects near these MMRs.

The paper is divided as follows. In Section \ref{s1}, we present the results of a many-particles numerical integration of the equations of motion of the major planets and a disk of planetesimals following the Nice model. This integration is used to determine the migration speed used in the following sections and also to present an example of the mechanism of creating fossilized detached objects that will motivate the rest of the work. In Section \ref{s2} we present a semi-analytic study of the Kozai 
mechanism inside the 2:5 and 1:3 MMRs. Section \ref{s3} shows the main results of our numerical
experiments without Neptune's residual migration. We devised a pathway for the 
formation of detached objects from primarily scattered particles, and relate it to the semi-analytic
study developed in Section \ref{s2}. Section \ref{rm} shows the main results of our 
numerical experiments {\it with} Neptune's residual migration.  We present a calculation of 
the ratio between detached objects with high and moderate perihelia and an estimate of the mass deposited near those resonances. Finally, we summarize 
our main results and give the conclusions in Section \ref{s5}.

\section {Results from a numerical integration with the Nice model}
\label{s1}

\begin{figure}
\resizebox{\hsize}{!}{\includegraphics{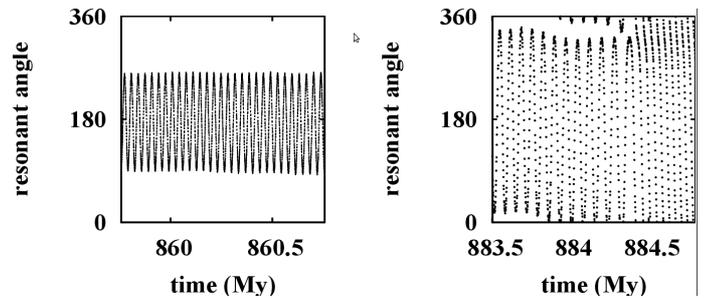}}
\caption{ Evolution of semimajor axes of the giant planets for a Nice model simulation. Each planet in this plot is identified by its present order of distance from the Sun.}
\label{r1}
\end{figure}

\begin{figure}
\resizebox{\hsize}{!}{\includegraphics{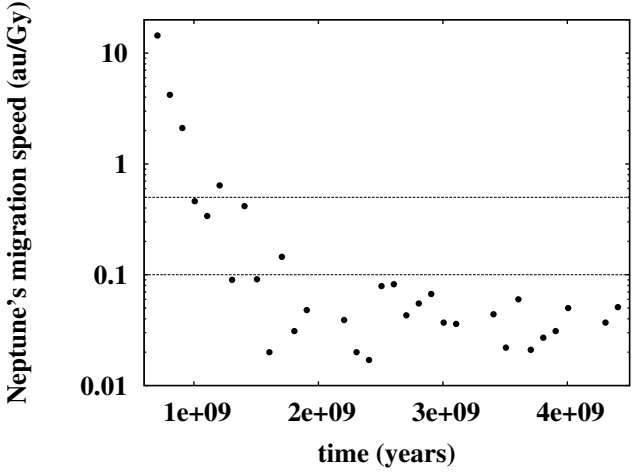}}
\caption{ Evolution of Neptune's migration speed during the Nice model simulation shown in Figure \ref{r1}. Horizontal lines stand for migration speeds $0.1au/Gy$ and $0.5au/Gy$}
\label{r2}
\end{figure}

 In this section we consider a numerical integration of the equations of motion of the planets and massive particles according to the Nice model \citep {tsiganis2005}. The initial orbital semimajor axes for the planets are $5.45au$, $8.18au$, $11.5au$ and
$14.2au$ for Jupiter, Saturn, Uranus, and Neptune, respectively. Their eccentricities are initially zero, and their inclinations are
either $0^{\circ}$ or $0.5^{\circ}$. The disk is initially composed of $10^3$ planetesimals with a total mass of $35$ Earth masses, and with a surface
density scaling as $r^{-1}$ and situated between $16.0$ and $40.0au$. They have zero eccentricities and inclinations. When the planets had stopped their mutual encounters, 
we cloned each planetesimal
$30$ times. Figure \ref{r1} shows the evolution of the semimajor axes of the planets just before and after the instability phase triggered by the close encounters between the planets. We focus particularly on Neptune's migration. Figure \ref{r2} shows the evolution of Neptune's migration speed. This is done by averaging Neptune's semimajor axes at every My, then averaging the migration speed taken from the previous averaged semimajor axes for every $100$ My. We notice that from $t=0.75$ Gy to $t=1.5$ Gy, Neptune's migration speed varies roughly from $5au/Gy$ to $0.1au/Gy$. This is the range of speeds used in the simulations in Section 5, which corresponds to Neptune's migration speed during a time span of roughly $0.85$ Gy. It must also be noticed that although after around $1.5$ Gy in the integration shown in Figure \ref{r2}, Neptune has an average migration speed below $0.1au/Gy$ if we do not perform an average of these speeds. We notice the absolute value of Neptune's migration speed above $0.1au/Gy$ during
most of the time ($78$\%). We might thus argue that in a sense Neptune has experienced a residual migration until today.

\begin{figure*}
\sidecaption
\includegraphics[width=12cm]{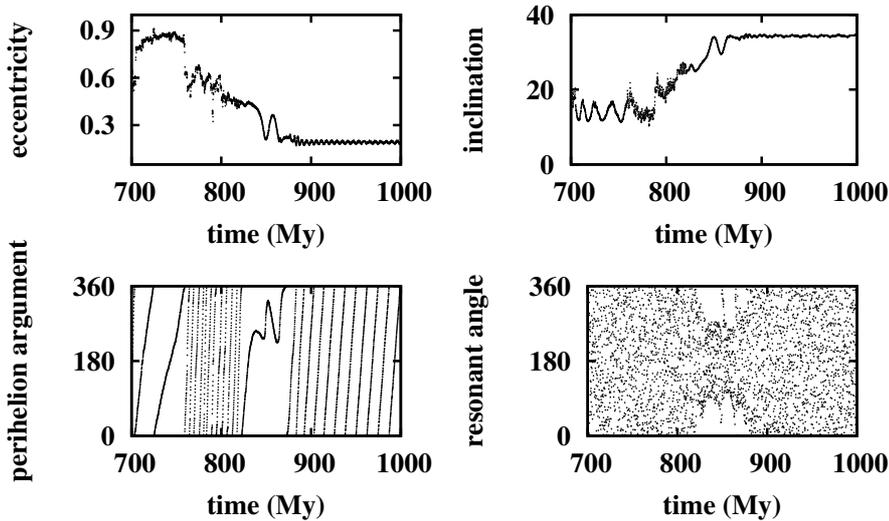}
\caption{Orbital evolution of a SDO that get trapped into the 2:5 MMR with Neptune, experiences the Kozai mechanism and eventually escaped the resonance being fossilized as a detached object near the 2:5 resonance. Angles are in degrees.}
\label{r3}
\end{figure*}

\begin{figure}
\resizebox{\hsize}{!}{\includegraphics{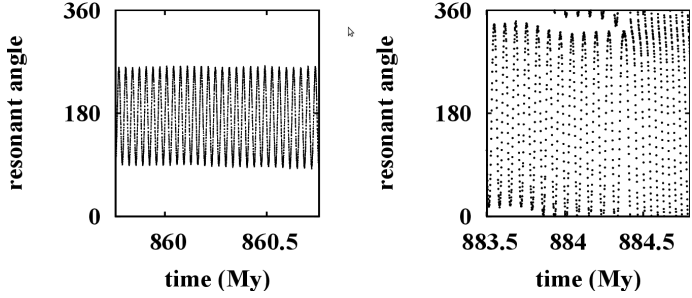}}
\caption{Evolution of the 2:5 resonant angle for the same particle as in Figure \ref{r3} at two different times giving more detail on its variation. In the left panel, the particle is well inside the resonance. In the right panel the libration amplitude of the resonant angle is very large and eventually it switches to circulation. That would roughly define the time of escape from the resonance. Angles are in degrees.}
\label{r4}
\end{figure}

This simulation also showed one instance of a trapping and later escape with respect to both the 2:5 and 1:3 resonances. Figure \ref{r3} shows such a case for the 2:5 resonance. We notice that a little after $0.8$ Gy to a little before $0.9$ Gy, the particle experiences the 2:5 MMR coupled with Kozai resonance. Somewhere before $0.9$ Gy, the particle escapes both resonances and gets apparently fossilized with fairly low eccentricity and high inclination. It is also instructive to follow the evolution of the resonant angle with more detail at two especific times of the particle's evolution. This is shown in Figure \ref{r4} where in the lefthand panel the resonant angle is librating with $\sim 90^{\circ}$ amplitude around $180^{\circ}$. At around $0.884$ Gy (right panel), the particle's resonant angle libation amplitude is very large and eventually turns to circulation, defining a escape from the resonance. 

\begin{figure}
\resizebox{\hsize}{!}{\includegraphics{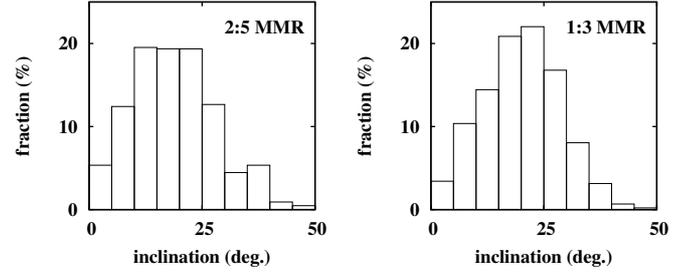}}
\caption{Distribution of inclinations of particles around the 2:5 and 1:3 resonances. These inclinations were taken from a time range from $0.7$ to $1$ Gy in the Nice model simulation shown in Figure\ref{r1}. }
\label{r7}
\end{figure}

Figure \ref{r7} shows the distribution of the particle's inclination around the 2:5 and 1:3 MMR, including a range of semimajor axes plus or minus $1au$ around the resonance's nominal semimajor axis. The perihelion distances are from Neptune's semimajor axis up to $5au$ beyond. These distributions will be useful for comparing with the simulations undertaken in the following sections. The data are from a time range between $0.7$ and $1$ Gy.

\section{A semi-analytic approach of the Kozai mechanism inside 2:5 and 1:3 MMRs}

\label{s2}

Based on the work of \citet{thomas1996}, \citet{gomes2005b}, \citet{gomes2011}, and \citet{gallardo2012}, we developed 
our study for 
particles trapped in the 2:5 and 1:3 MMRs with Neptune. We consider the four major planets
 in planar and circular 
orbits. Our aim in this section is to find an average perturbation of the planets on 
the particle. This amounts to computing an average of the Hamiltonian in the mean longitudes 
of the planets and particle. When the particle experiences a MMR, the problem can be 
reduced to 
the calculation of a single integral, since we have a fixed relation between the mean 
longitudes
 of the planet and the particle given by the resonant angle (when its libration 
amplitude is zero). The general resonant angle is defined by
\begin{equation}
 \qquad\qquad\qquad \phi=(p+q)\lambda_N-p\lambda-k_1\varpi-k_2\Omega
\label{resang}
\end{equation}
where $\lambda_N$ and $\lambda$ are the mean longitudes of Neptune and the particle, the integers $p$ and $q$ are
 the degree and order of the resonance, $k_1,k_2$ are integers satisfying $k_1+k_2=q$, $\varpi$ is the particle's longitude of perihelion and $\Omega$ its longitude of the ascending node. In the
rest of the paper, we always consider $\phi_{2:5}$ and $\phi_{1:3}$ as the
eccentricity-type ones, those that do not include the longitude
of the node, defined by
$\phi_{2:5}=5\lambda-2\lambda_N-3\varpi$ and $\phi_{1:3}=3\lambda-\lambda_N-2\varpi$.

In real N-body dynamics the particle always presents a finite libration amplitude of the resonant angle. 
Thus we turn back again to a double integral where the mean longitudes are still related through the 
resonant angle, which is made to vary according to a sinusoid of the given amplitude. Owing 
to the averaging in the particle's mean longitude, its canonical conjugate 
\begin{equation}
 \qquad\qquad\qquad \qquad\qquad L=\sqrt{\mu a}
\label{eqL}
\end{equation}
is constant. In Eq. (\ref{eqL}), $a$ is the semimajor axis of the trans-Neptunian object and $\mu$ the gravitational 
constant times the Sun's mass. Due to rotational 
invariance in the longitude of the node, its conjugate (divided by $L$) 
\begin{equation}
 \qquad\qquad\qquad\qquad H=\sqrt{1-e^2} \cos(i)
\label{eqH}
\end{equation}
will also be a constant, where $i$ is the particle's orbital inclination and $e$ its eccentricity. In the end,
the averaged Hamiltonian will be in the form $E=E(G,g)$ where the canonical conjugated pair represents
$g\equiv \omega$ (argument of perihelion), and $G=\sqrt{\mu a(1-e^2)}$. Therefore, one has two constants of motion: the Hamiltonian ($E$), which is related to the energy, and $H$, which stands for the particle's angular momentum projected on the reference (planets' orbital) plane. Since our
Hamiltonian is one degree of freedom, it is possible to draw level curves for fixed values 
of $E$ and $H$.

As pointed out by \citet{kozai1962}, when the orbital inclination of the asteroid is high enough 
($39^{\circ}\leq i\leq 141^{\circ}$ for asteroids in the main belt), its argument of perihelion starts to 
librate around fixed values ($\omega=k.90^{\circ}$, with $k$ an integer), and the pair ($e,i$) will be 
related by the constant of motion $H$. This kind of behavior has been known as Kozai resonance ever since
 then. Other authors have shown that if a TNO is trapped in a MMR with Neptune, it can present the 
Kozai dynamics for inclinations lower than $39^{\circ}$ \citep{gomes2003a, gomes2003b, gomes2005b,
 gomes2011, gallardo2012}. Moreover, if the resonance is of type $1:N$, the classical 
Kozai
centers vanish and new asymmetric centers ($\omega\ne k.90^{\circ}$) take their place 
\citep{kozai1985, gallardo2012}. This is because the centers of 
$\phi_{1:N}$ are different from $0^{\circ}$ and $180^{\circ}$, and new terms in the mean 
resonant disturbing function arise \citep{gallardo2012}.

\begin{figure}
\resizebox{\hsize}{!}{\includegraphics{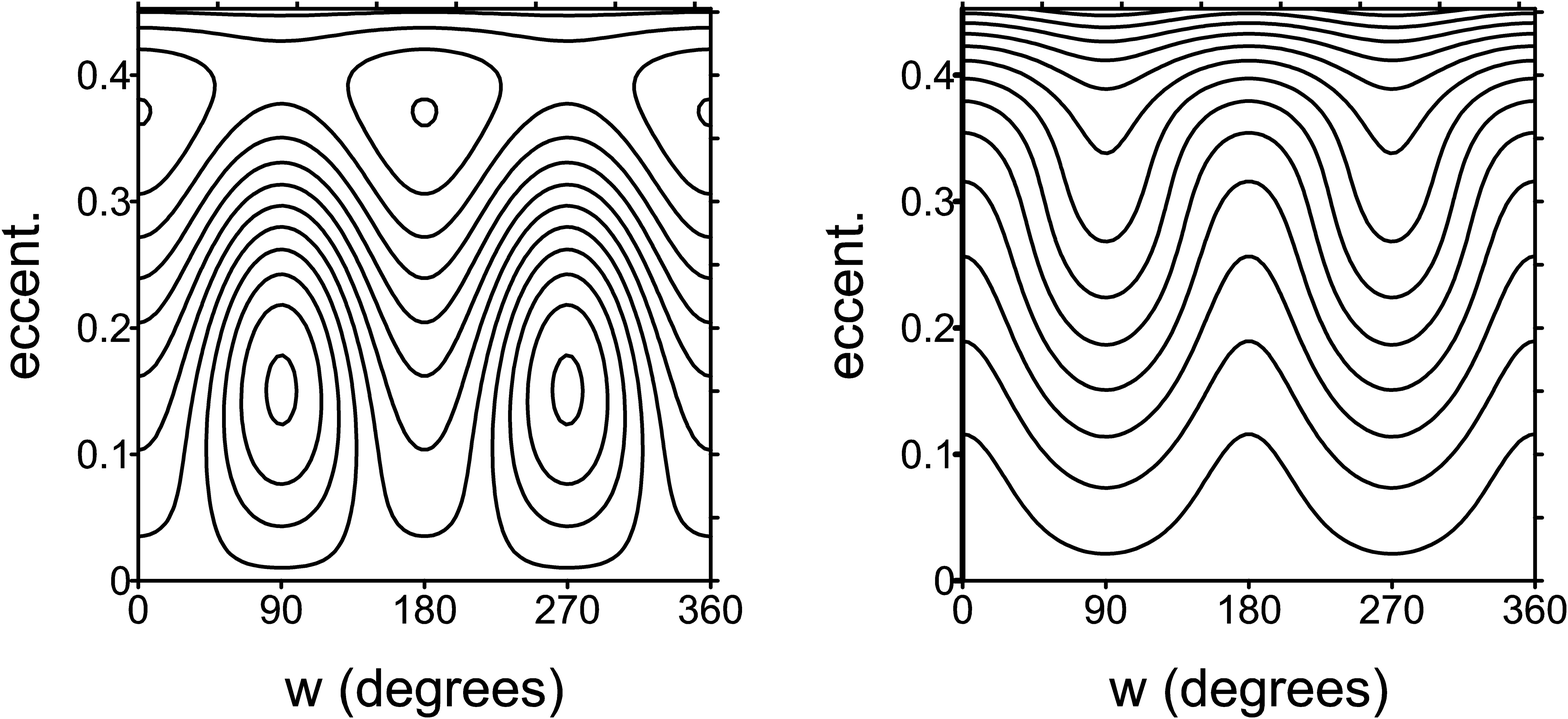}}
 \caption{(a) Energy level curves {\it e vs.} $\omega$ for a particle trapped in the 2:5 MMR, 
librating around $\bar\phi_{2:5}=180^{\circ}$ with $70^{\circ}$ of amplitude. We notice the appearance of the 
classical Kozai centers at $\omega=k.90^{\circ}$. The particle can experience large variations in 
the eccentricity and inclination.  In (b) we have the same initial conditions as (a), 
except for the libration amplitude thas is now equal to $120^{\circ}$. We notice the 
disappearance of the 
classical Kozai centers at $\omega=k.90^{\circ}$. The large ($e,i$) variations are not possible,
 so each orbit will either have high eccentricity and low inclination or vice-versa.}
 \label{f1}
\end{figure}

From the semi-analytical approach described above we
can build Figure \ref{f1}, which shows two examples of diagrams {\it e vs.} $\omega$ 
for particles trapped
 in the 2:5 MMR. In figure \ref{f1}(a) the particle has $H=0.851$, $a=55.47au$, 
$\bar\phi_{2:5}=180^{\circ}$, and $70^{\circ}$ of libration amplitude. Each level curve stands for a 
specific energy level and presents the classical
Kozai dynamics where $\omega$ librates around the centers $\omega=k.90^{\circ}$. Therefore it is 
expected that a particle with these caracteristics presents anti-phase large variations in 
the 
pair ($e,i$) due to the conservation of $H$ and libration  of $\omega$ around the classical 
centers. If the 
libration amplitude  of $\phi_{2:5}$ increases, the relation between the mean longitudes is 
almost lost, and the 
motion will be similar to non-MMR Kozai dynamics. Figure \ref{f1}(b) shows almost the same 
case as Figure \ref{f1}(a),
except that we now take $120^{\circ}$ for the libration amplitude of the resonant angle 
($\phi_{2:5}$). Since $H$ is still 
a constant of motion, the lower curves represent orbits that have low eccentricities and 
high inclinations,
 while the higher ones have lower inclinations and higher eccentricities. 

\begin{figure*}
\sidecaption
\includegraphics[width=12cm]{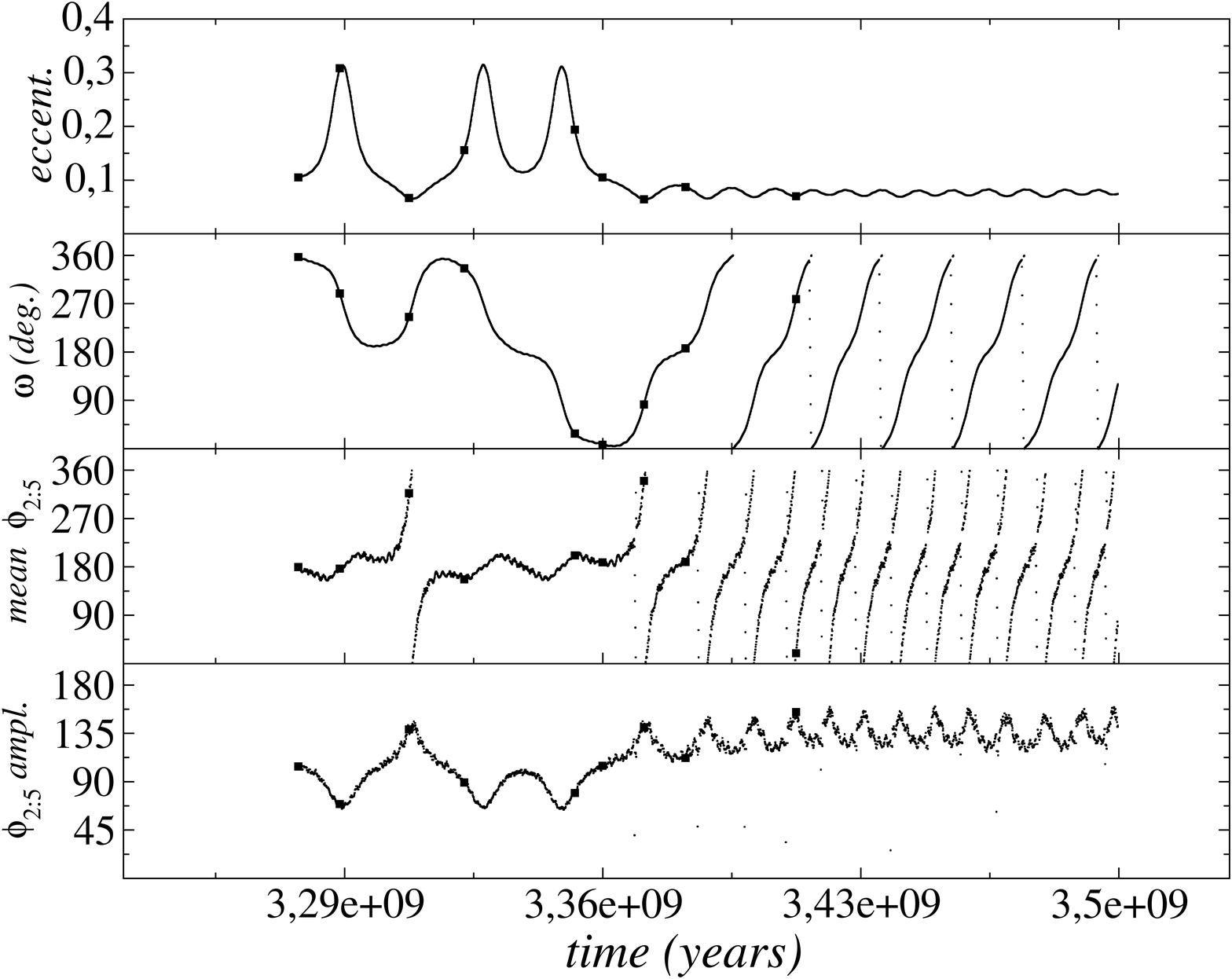}
\caption{Results of a numerical simulation in which the particle experiences the 2:5 MMR+Kozai, 
and the disappearance of both due to a large libration amplitude of $\phi_{2:5}$. The final orbit has 
low eccentricity, hence high inclination (not shown), because it is in a state called 
hibernating mode, since it can turn out to experience the MMR+Kozai due the chaotic behavior presented. The
four panels show the time evolution of the eccentricity, argument of perihelion, mean resonant 
angle ($\phi_{2:5}$), and its amplitude, respectively. Angles are in degrees. The plotted squares stand for specific times for which energy level curves are plotted in Fig. \ref{f4}.}
\label{f3}
\end{figure*}

\begin{figure}
\resizebox{\hsize}{!}{\includegraphics{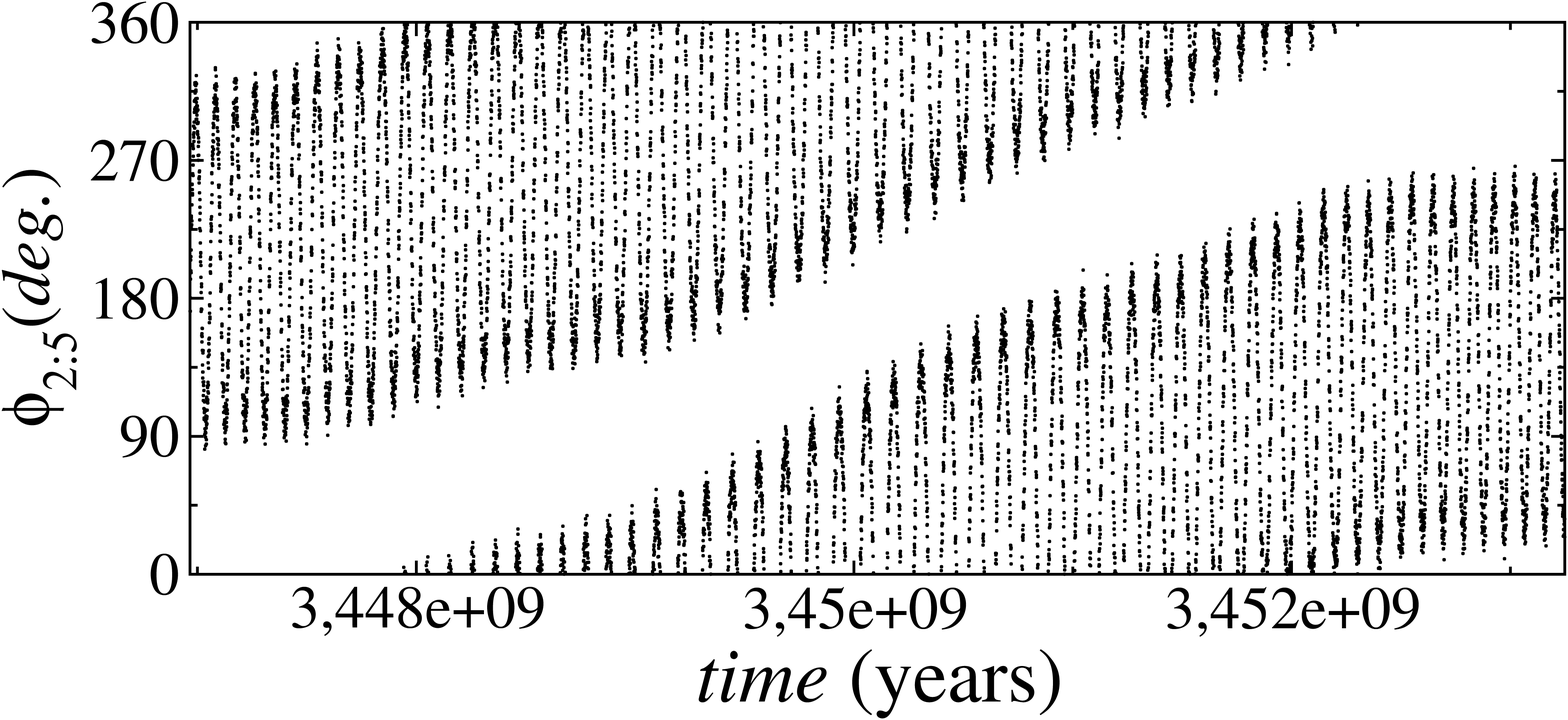}}
\caption{The variation in the same resonant angle as in Figure \ref{f3} not averaged and zoomed in time so as to make the libration period timescale apparent. This behavior is common to exterior MMRs with Neptune coupled with Kozai resonance. Even when the particle is not in the hibernating mode, this behavior appears as, for instance, during the time between $3.3$ Gy and $3.32$ Gy as shown in Figure \ref{f3}.}
\label{deta-hib}
\end{figure}

In Figure
\ref{f3} we present a narrow window in time of one of our numerical 
simulations in which 
the process described above appears for a particle whose $H=0.851$. It experiences the 2:5 
MMR coupled with the Kozai resonance for $\sim 100My$ ($\sim 3.27Gy$ to $\sim 3.37Gy$). 
Then it accesses 
a dynamical mode with low eccentricity or high perihelion distance (and high inclination) 
and high libration amplitude of the MMR angle for the rest of the simulation. We call this last dynamical mode 
of {\it hibernating mode} because the particle may ``awake'' at any time and return to  the
 MMR+Kozai dynamics. In Figure \ref{deta-hib}  we show a zoom in time of Figure \ref{f3}, for the unaveraged resonant angle. Although this behavior might be mistaken for a circulation, Figure \ref{deta-hib} shows that the resonant angle is really librating when the libration period timescale is considered. On the other hand, the libration center is not constant but circulates. This is why it is more instructive to show the variation in the resonant angle in Figure \ref{f3} as averages in libration amplitude and libration center. It is important to note that in these exterior MMRs coupled with Kozai resonance, the variation of the libration center is the rule and not the exception.

\begin{figure*}
\centering
\includegraphics[width=17cm]{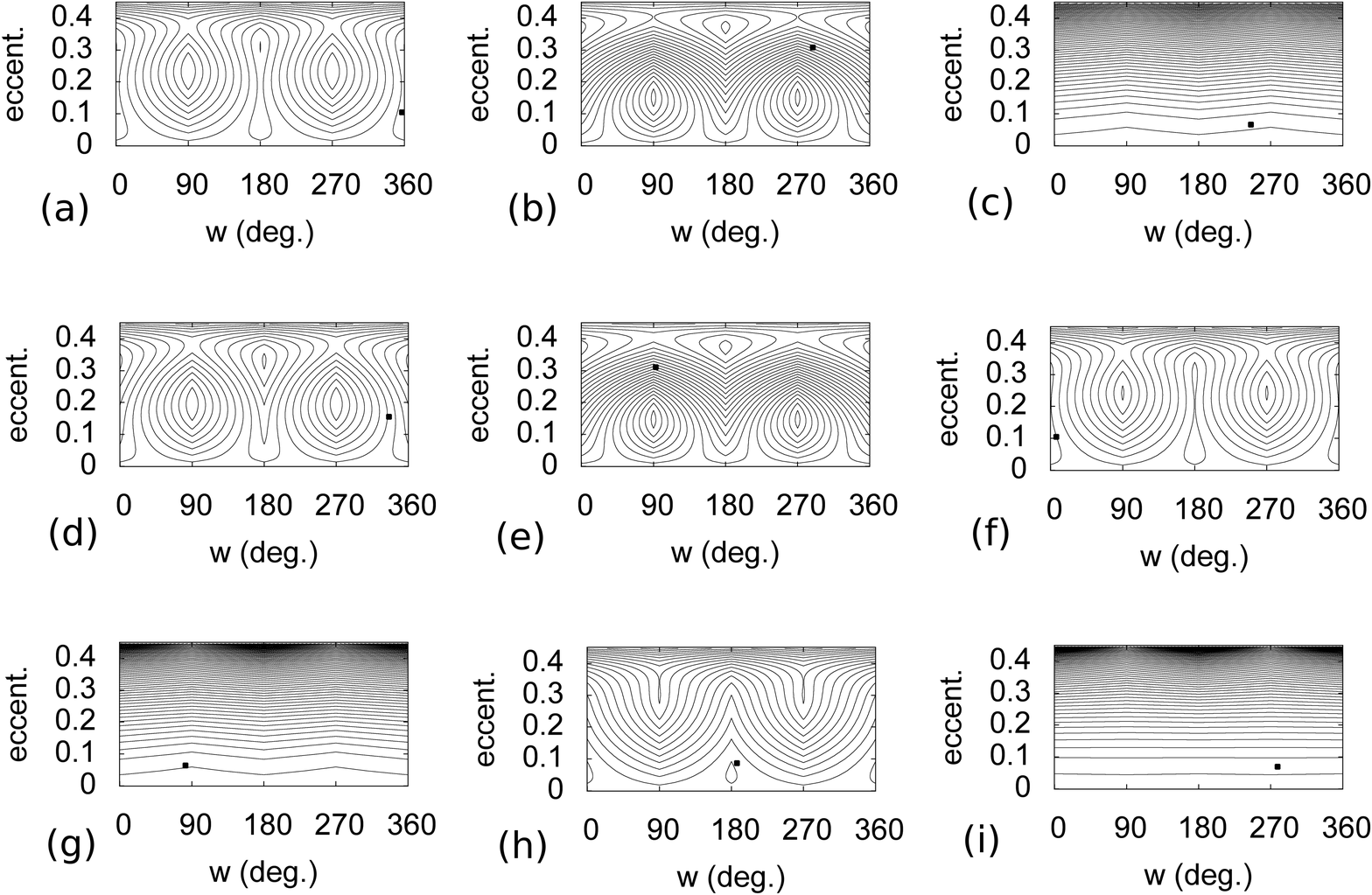}
\caption{This sequence of shots are associated to the sequence of points in Figure \ref{f3} represented as squares. We plot $e$ vs. $\omega$ of the particle also as squares. At each point, the amplitude and libration center of the resonant angle enable us to plot the energy level curves associated with each square, so the squares follow equipotential curves in time-varying energy level diagrams. The main point is to show the random character of the entrance into the hibernating mode. Panels c and g are very similar and correspond to peaks in libration amplitude as shown in Figure \ref{f3}. The transition to panels d and h shows the subtle difference between entering and not entering the hibernating mode.}
\label{f4}
\end{figure*}

Figure \ref{f4} presents the  energy level curves 
associated with each square in 
Figure \ref{f3}. The process of entering a hibernating mode seems quite subtle. In panel 
(c), the particle had a high libration amplitude, and the  energy  level curves were flat 
like in a non-MMR mode. In panel (d), the libration amplitude had already decreased enough 
 to lead to the appearance of closed curves ($\omega$-librating) with large variations 
in the eccentricity. The particle thus manages to escape the hibernating mode along one of 
these Kozai curves. A similar process is observed from panels (g) to (h), but in this case 
the decrease of the libration amplitude was not large enough to create an ascending path 
for the particle, and as shown in panel (i), it enters the hibernating mode. The exact 
circumstances that enable a particle to enter or not the hibernating mode seems quite 
complicated and is most likely suitable to being studied as a probabilistic event, although 
that is beyond the scope of this paper. Owing the chaotic behavior of this dynamics, a 
particle once in hibernating mode can again be trapped  in the MMR coupled with Kozai. To place the particle in a fossilized
orbit, without the risk of being trapped again into the MMR, a complementar mechanism is needed. This 
will be discussed in the next sections.

\begin{figure}
  \centerline{\includegraphics[width=0.999\linewidth]{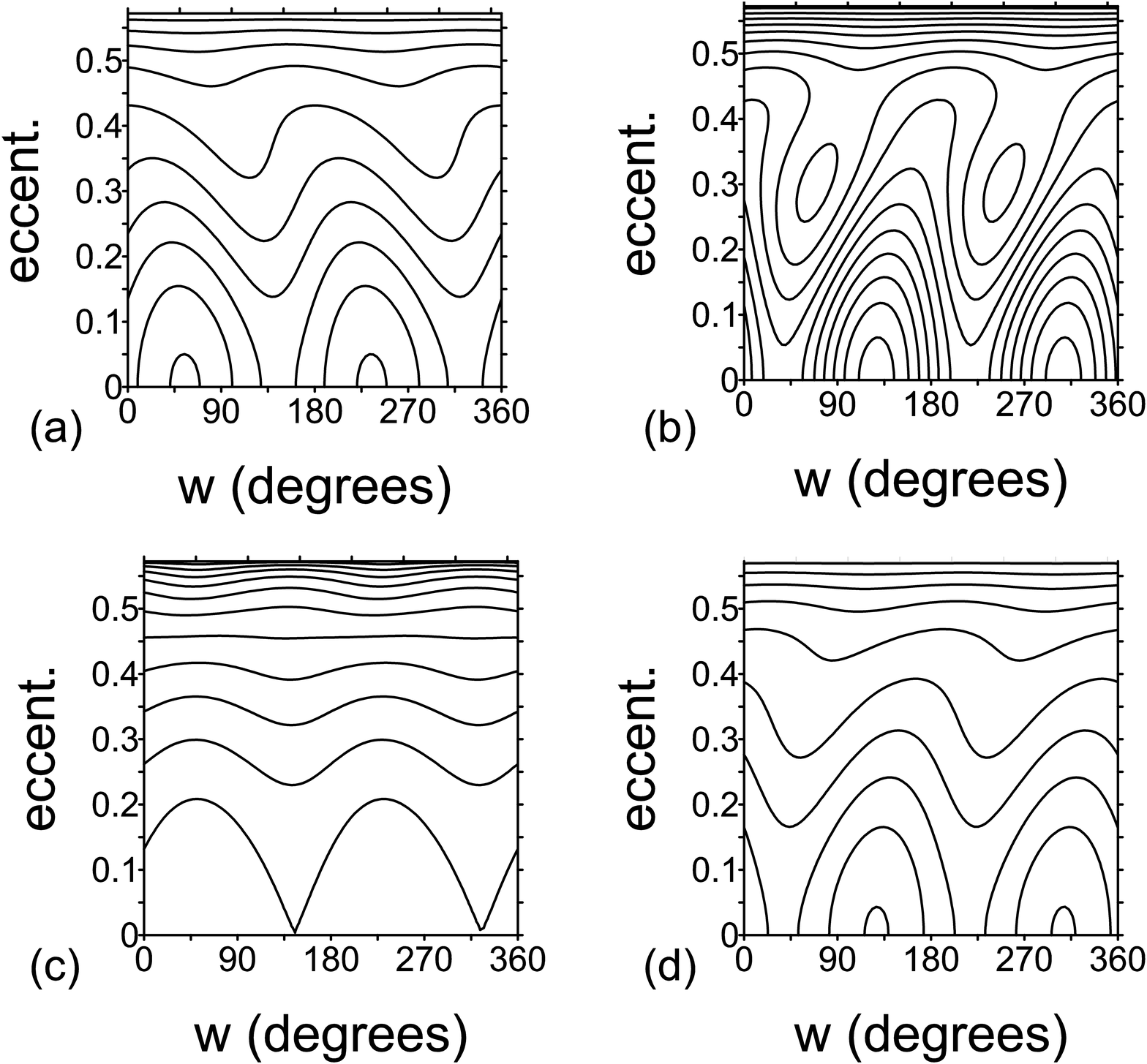}}
 \caption{Energy level curves {\it e vs.} $\omega$ for the 1:3 MMR. In (a) $\bar\phi_{1:3}=70^{\circ}$ with $30^{\circ}$ of amplitude,
in (b) $\bar\phi_{1:3}=285^{\circ}$ with $5^{\circ}$ of amplitude, in (c) $\bar\phi_{1:3}=70^{\circ}$ with $60^{\circ}$ of amplitude,
and in (d) $\bar\phi_{1:3}=285^{\circ}$ with $30^{\circ}$ of amplitude.  The asymmetric centers of the Kozai 
resonance are clear in (b), and the flattened curves in (c) indicate that the hibernating mode 
can also be reached. All cases have $a=62.63au$ and $H=0.82$.}
\label{f5}
\end{figure}

For the 1:3 MMR, the same main features may occur, except that in this case when the inclination is 
high enough the asymmetric centers of the Kozai resonance appear. Figure \ref{f5} presents
these asymmetric centers for different values of the libration center $\bar\phi_{1:3}$ and libration amplitudes. In all 
cases the particle preserves $a=62.63au$ and $H=0.82$. Again, if the amplitude of $\phi$ is large, 
the flattened curves appear, indicating that the hibernating mode is also possible for this resonance.

%
\section{Numerical simulations without migration}
\label{s3}

We present the results of seven numerical simulations of the equations of motion of the four giant 
planets in their current orbits referred to the ecliptic plane at Julian date 2454200.5 and a total of 22,500 massless scattered particles. The integrations were 
performed for $4.5Gy$ in steps of 0.5 years, which is approximately equal 
to 1/20 of 
the smallest orbital period (Jupiter). We used the hybrid integrator of the Mercury package \citep{chambers1999}
for these numerical simulations.
The particles were divided into two groups around the mean motion resonances 2:5 ($a_{2:5}=55.5164au$)
 and 1:3 ($a_{1:3}=62.6912au$) with Neptune. Initial conditions
 were assigned randomly, respecting the following limits: 
$54.5161\leq a\leq 56.5161au$ or  $61.6912\leq a\leq 63.6912au$ for semimajor axis, 
 $30< q \leq 35au$ for perihelion distance, $0\leq i\leq 50^{\circ}$ for inclination, $[0,360^{\circ}]$
 for the angles argument of perihelion $\omega$, longitude of the ascending node $\Omega$, and
 mean anomaly $l$. The planets interact among themselves and disturb the massless particles, 
which are discarded from the simulations
 when $ a <52.50au$  or $ a>65.75au$, decreasing the large cpu time spent to 
complete this kind of simulation. Since the particles are considered to be massless, there is no exchange of energy and 
angular momentum between the disks and the planets, and therefore there is no planetary migration.

\begin{figure*}
\sidecaption
\includegraphics[width=12cm]{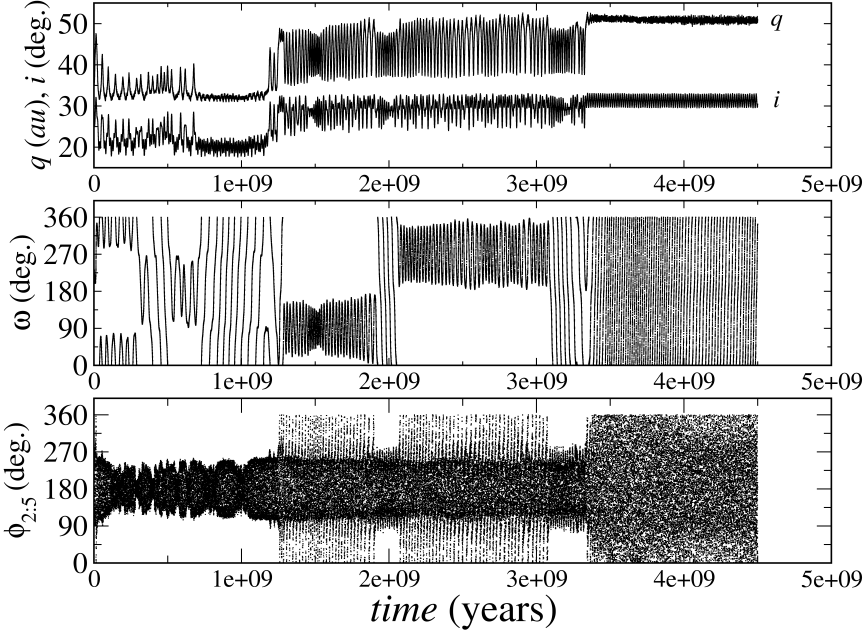}
\caption{Time evolution of a particle captured by the 2:5 MMR with Neptune. Upper panel shows the evolution
of the perihelion distance, {\it q}, and inclination, {\it i}. The middle panel shows the evolution of
the argument of perihelion, $\omega$, and the bottom panel presents the resonant angle 
$\phi_{2:5}$. We notice several dynamical modes like the Kozai+MMR, and the hibernating
 mode. In the end the particle is in a non-fossilized detached orbit.}
\label{f7}
\end{figure*}

\begin{figure*}
\sidecaption
\includegraphics[width=12cm]{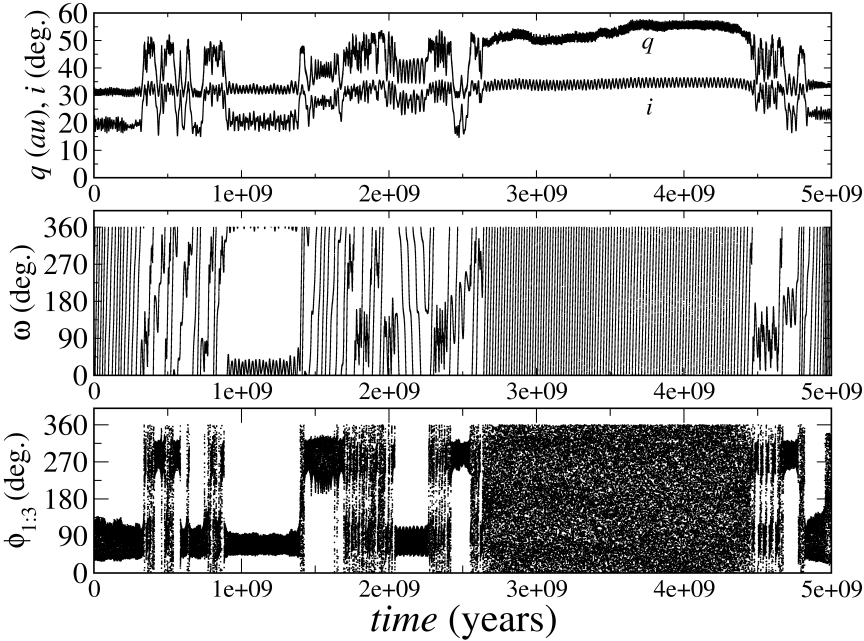}
 \caption{ Time evolution of a particle captured in the 1:3 MMR with Neptune. Upper panel
 shows the evolution
of the perihelion distance, {\it q}, and inclination, {\it i}. The middle panel shows the evolution of
the argument of perihelion, $\omega$, and the bottom panel presents the resonant angle 
$\phi_{1:3}$. It is possible to note one of the asymmetric centers of the Kozai resonance between
$800My$ and $1.4Gy$. Despite the particle 
remains in the hibernating mode for a long time, the resonances restart in the Kozai mode and the large perihelion
variation dynamics are installed at the end of the simulation.}
 \label{f8}
\end{figure*}

We now present two representative figures that show the most interesting dynamic modes found. 
Figure \ref{f7} shows the case of a particle initially in the region of the 2:5 MMR,
 and in Figure \ref{f8} the particle was initially near the 1:3 MMR. The particle in Figure \ref{f7} starts
 in Kozai resonance and MMR, 
as seen in the middle and bottom plots and the coupled oscillation between the 
perihelion distance and inclination in the top plot. This Kozai mode, whose centers are $0$ and 
$180^{\circ}$, lasts about 700My.  After that, the argument of perihelion starts to circulate with a long 
period, and the particle remains in a high-eccentricity orbit, whose perihelion 
is close to Neptune's orbit, basically following similar dynamics to one of the top flattened curves of Figure 
\ref{f1}-b, for $\sim400My$. Close encounters
with Neptune do not occur because it remains in MMR, as seen by the 
libration of $\phi_{2:5}$. Between $1.2Gy<t<3.4Gy$, the particle experiences a new type of Kozai mode,
 where the oscillations of eccentricity and inclination have greater amplitude and the libration centers
 of $\omega$ are most of the time at $90^{\circ}$ and $270^{\circ}$, but it circulates during  some time around $2$ Gy. This behavior is also consistent with that shown in Figure
\ref{f1}-a. Although it appears that the angle $\phi_{2:5}$ does not librate in this interval, with
 a zoomed plot similar to Figure \ref{deta-hib}, one notices that the center mostly librates around $180^{\circ}$ and 
undergoes some rapid changes, 
which on a scale like the one 
presented in Figure \ref{f7} leads to a misinterpretation of circulation. For $t > 3 Gy$, the argument of the perihelion starts to circulate again before the particle reaches a mode with low eccentricity 
(high perihelion 
distance) and high inclination to the end of the numerical simulation (hibernating mode).

A 
similar analysis can be made ​​of Figure \ref{f8}; however, it is important to note 
that
the rise of the asymmetric centers of the Kozai resonance, as 
noted between 900My and 1.4Gy and in the energy level curves presented in Figure \ref{f5}-b. Another noteworthy aspect is that despite spending a long period in 
hibernating mode, the resonances may return to be active and bring the particle back to the resonant dynamics
where large perihelion variations are possible. This is noticed in the last $500My$
 of the simulation in 
Figure \ref{f8}. 

 Up to now, we have shown that  
 detached objects can be
 formed in a solar system without planetary migration.  
But it is not guaranteed that they will remain as detached or return to the resonant dynamics. 
However, no matter how slowly, the planets are always migrating, since they interact with
the TNOs and with the comets from the Oort cloud. In this sense, it is possible currently to form 
detached objects and fossilize their orbits. This subject is addressed 
in the next section.

\section{Imposing a residual migration on Neptune}
\label{rm}

We now proceed to a new set of numerical integrations with no migration. This time Jupiter, Saturn, and Uranus are at
their current positions, as in the last section, while the mean semimajor axis of Neptune was decreased to 
$\bar{a}_N=29.8au$.
A total of 16,000 particles were initially considered. The limits on the particles' initial 
conditions and other 
integration parameters are equivalent to those in section \ref{s3}, except that the 
semimajor axes of the particles are distributed around $a=54.8911\pm1au$ and $a=61.9854\pm1au$. 
This new set of integrations was done to check whether the
main characteristics of the first set of integrations were maintained and to introduce the 
possibility of including a residual
migration on Neptune for further analysis with migration. One finds that the global features are 
the same for the two sets of numerical simulations. Similar results are also
possible in the case, as shown by Figures \ref{f7} 
and \ref{f8}, where Neptune is placed before its actual position. One also finds that $12\%$ of
the particles attain $q>40au$ regardless of whether they survive or not for the age of the solar system, 
but only one fourth of them reach hibernating mode. In these integrations, we identify the hibernating mode whenever $q>40au$
for $\Delta t\ge100My$.  With this condition, one can identify, in a general way, the hibernating cases 
inside the large data set generated by the simulations without looking one by one. 
Since in the hibernating mode the 
resonant angle circulates or librates with
very large amplitude, these particles become potential candidates to fossilization while Neptune 
migrates.

With these simulations we want to understand 
 the behavior of the particles when Neptune is still migrating to its current position. To
accomplish this task we artificially added a force that depends on the velocity components into Neptune's equations
of motion in order to linearly increase its semimajor axis. Our idea was to study the final phase of a planetary migration. We considered that Neptune's migration speed  comes from the Nice model integration described in Section \ref{s1}. We also note that for a smooth migration, roughly the same migration speeds are obtained. \footnote{We performed a basic numerical integration with the four giant planets initially placed so as to experience a smooth migration and a disk of massive planetesimals with the same total mass as in the Nice model simulation described above, and we noticed Neptune's migration speed above $0.1au/Gy$ during most of the time up to $4.5$ Gy.} This result argues for the robustness of the mechanism presented here. We only consider Neptune's migration since the farther planet migrates much faster than the others, and dynamical processes experienced by TNOs (like mean motion resonances) are  almost solely related to Neptune. 

The first experiment consisted of choosing two different points in the orbital evolution of a particle.
For one of these points, the particle is experiencing the Kozai dynamics with strong variations in the 
eccentricity and the inclination, and the particle is in the hibernating mode for the other point. 
In this experiment we only considered that particle and the planets at those two chosen 
times. For either point (particle and time), new integrations were started as many times as the chosen migration speeds
for Neptune in the interval $[0.1 ; 5.0]au/Gy$, with steps of $0.01au/Gy$. Therefore, the resonant semimajor axis also 
migrates. The simulations were stopped whenever Neptune reaches its current average  semimajor axis of $30.11au$. 

\begin{figure}
\resizebox{\hsize}{!}{\includegraphics{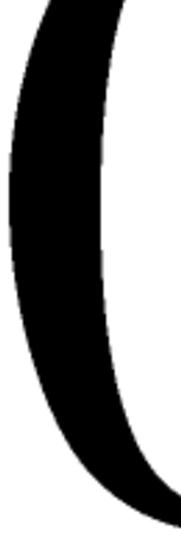}}
 \caption{ Diagram $V_{mig}$ {\it vs.} averaged semimajor axis for particles that were initially in 2:5 MMR 
with Neptune. The average is calculated over the past $5\%$ of each simulation. The black filled circles
represent particles that started the new simulations in the hibernating mode, while the open circles
represent those that started in the Kozai mode. The vertical lines are the approximate resonant
semimajor axis before (at left) and after (at right) the residual migration of Neptune. It is clear
that particles initially in the hibernating mode tend to escape the MMR, becoming fossilized objects, 
while the ones started in the Kozai mode, in the opposite way, tend to stay trapped in the MMR and 
follow the migration.}
 \label{f9}
\end{figure}

\begin{figure}
\resizebox{\hsize}{!}{\includegraphics{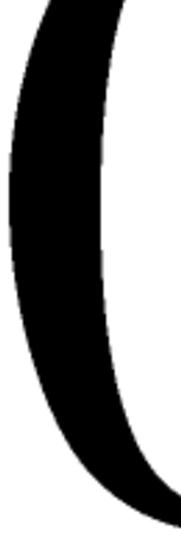}}
 \caption{ Analogous to the Figure \ref{f9} for particles that were initially close to
the 1:3 MMR with Neptune.}
 \label{f10}
\end{figure}

By this 
experiment it is possible to check the particle's ability to escape from each dynamical mode and 
get fossilized in a smaller semimajor axis than the resonant ones. It is also possible to verify whether 
such escapes are related to the migration speed of Neptune. Figures \ref{f9} and \ref{f10} show
the results of Neptune's migration rates as a function
of the averaged semimajor axis over the final $5\%$ of each integration for particles that were
initially in the regions of the 2:5 MMR and 1:3 MMR, respectively. They also show
the approximate values of the resonant semimajor axis at the beginning and end of Neptune's residual migration \footnote{it is interesting to note that particles that escape the MMR immediately after the beginning of the migration are located at an average semimajor axis below the resonant one. This is because there is a discontinuity between average resonant semimajor axis and average nonresonant semimajor axis at the border of the resonance}.
It is noted that in almost $100\%$ of the cases initially in Kozai dynamics, the particles
tend to follow Neptune's migration, and continue captured in the MMRs. 

There are a few cases where
the particle started in the Kozai dynamics and escaped the MMR (Figure \ref{f10}). Analyzing 
the data from these simulations, one notices that, before the particle escaped the resonance, it switched to the hibernating mode. As for the case where the particles
begin in the hibernating mode (filled circles) one has the opposite situation, where most particles tend 
to escape the MMR easily, being fossilized without migrating. In the simulations where the particle 
was initially in the hibernating mode and followed Neptune's migration, we also observed that it 
switched the dynamic mode (from hibernating to Kozai) before starting the migration. Thus, we could claim through this experiment that to become a fossilized object with low eccentricity, the particle must reach the 
hibernating mode while Neptune is still migrating. The escapes from resonance at a specific point in the evolution were shown to be a function almost only of whether the particle was in Kozai or hibernating mode with barely any correlation 
with Neptune's migration rate in the studied range.

The experiments just performed suggest that escapes
from the hibernating mode is a near 100$\%$ chance event
for migration speeds above 0.1$au/Gy$, thus, as Figure \ref{r2} shows,
particles must escape from resonance in hibernating mode
until at least 3Gy before the present date. However, it is not obvious
that escapes after 1.5$Gy$ at hibernating mode are unlikely.
In fact, although the migration speed is on average below
0.1$au/Gy$ during the past 3$Gy$ of integration, we notice that
this speed can fluctuate between positive and negative values,
being often above 0.1$au/Gy$ in absolute value. So as to
check how likely escapes from resonance at hibernating
mode are for Neptune  migration speeds, such as those in the great integration
shown in Figure \ref{r2}, we performed the following experiment.
Everything were the same as for the experiments
that yielded Figures 13 and 14, except that the migration speed
imposed on Neptune will be dictated by the migration experienced
by Neptune in the great integration that yielded
Figure \ref{r2}. We thus computed Neptune's migration speed through
the differences in the average Neptune’s semimajor axis in
successive 5$My$ time ranges. Then we applied these successive
migration speeds, which was changed at every 5$My$, to
Neptune coming from the static integrations. We only used
the initial conditions at hibernating mode for both the 2:5
and 1:3 resonances as those used in the experiments that
yielded Figures \ref{f9} and \ref{f10}, in the beginning of this section. For
either resonance we did 25 numerical integrations imposing
Neptune’s migration speed as those from the great integration
as if Neptune started at 1.6$Gy$ until 4.0$Gy$ with steps of
0.1$Gy$. All integrations extend to 4.5$Gy$. Figure 15 shows the results
of this experiment, where we show Neptune's average
semimajor axis orbital evolution, and for each integration
started at the time shown in the figure, we plot ``E'' for an
escape event and ``R'' for a still in resonance event. We find
that there are still escapes from the hibernating mode during the 
last 3$Gy$ of Neptune's mild migration for a fraction between
one half and one third of the cases.

\begin{figure}
\resizebox{\hsize}{!}{\includegraphics{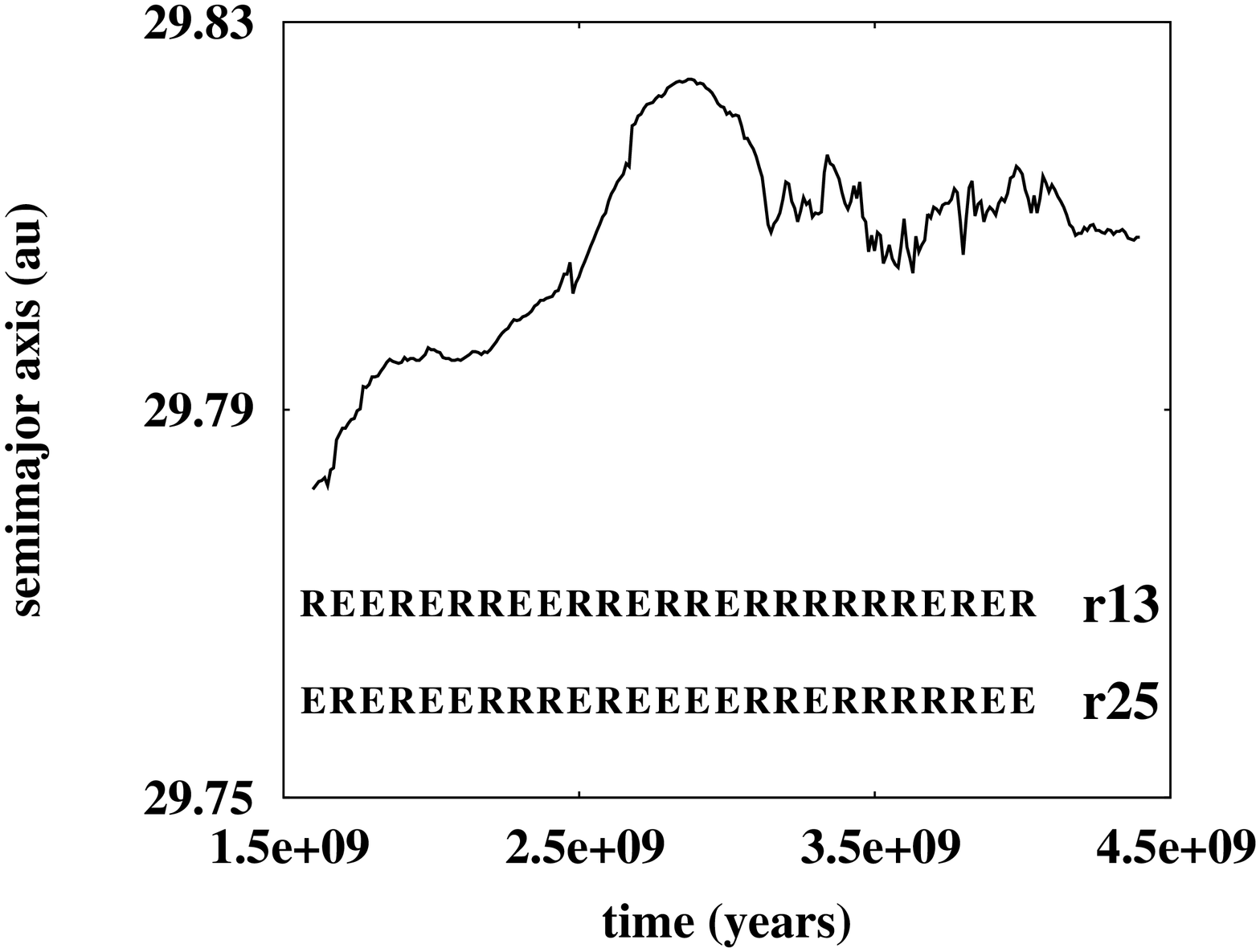}}
 \caption{Average evolution of Neptune’s semimajor axis with Neptune’s initial semimajor axis coming from
the static integrations and experiencing migration with the same migration rates as those suffered by Neptune
for the last 3Gy of the great integration shown in Figure 2. Also plotted is the result of the application of Neptune’s
migration on particles started at hibernating mode:``R'' stands for a particle that remained in resonance to
the end of the integration and ``E'' for a particle that eventually escaped the resonance before the end of the
integration. The time associated with either of these symbols stands for the initial time of integration with
Neptune experiencing the semimajor axis evolution shown in this plot. All integrations are extended to 4.5Gy}
 \label{f15-Rodney}
\end{figure}

\begin{figure}
\resizebox{\hsize}{!}{\includegraphics{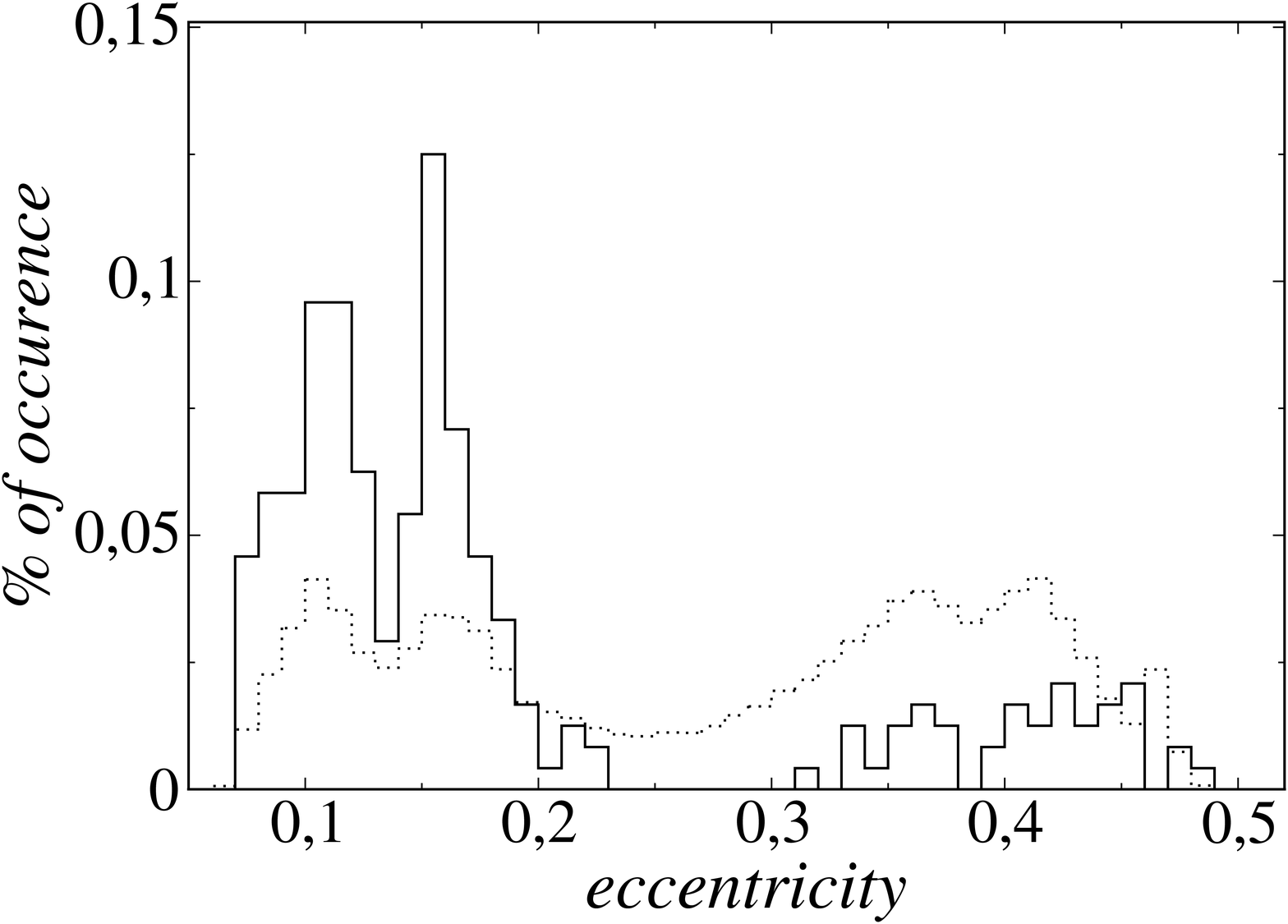}}
 \caption{ Comparison between the eccentricities distribution for the evolution of the 9 original particles
 in the 2:5 MMR region (dotted histogram) and the averaged eccentricities for the test particles 
that escaped the 
resonance while Neptune was migrating, becoming fossilized (continuous black histogram). Peaks at 
$e\sim0.12$ and $e\sim0.18$ suggest that $2004XR_{190}$-like orbits might be formed close to 2:5 
MMR with Neptune.}
 \label{f11}
\end{figure}

\begin{figure}
\resizebox{\hsize}{!}{\includegraphics{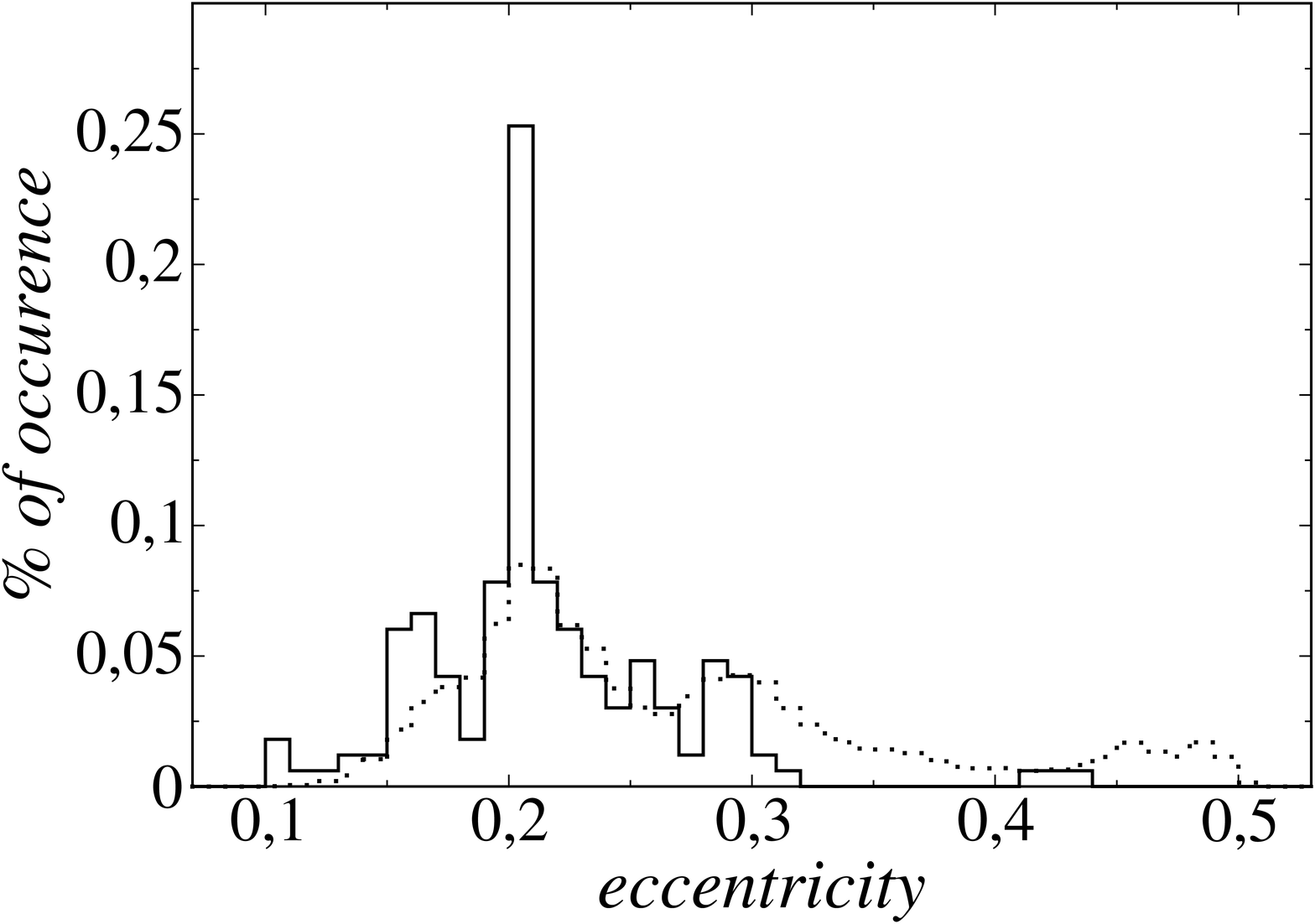}}
 \caption{ Comparison between the eccentricities distribution for the evolution of the 7 original particles
 in the 1:3 MMR region (dotted histogram) and the averaged eccentricities for the test particles 
that released the resonance while Neptune migrates, becoming fossilized (continuous black 
histogram). The peak at $e\sim0.21$ suggests that $2004XR_{190}$-like orbits might be formed 
close to 1:3 MMR with Neptune.}
 \label{f12}
\end{figure}

To find the most probable values of the eccentricities for fossilized particles, 
after they escape the MMR, we did a complementary experiment. This time we fixed 
the residual migration rate of Neptune in $V_{mig}=0.5au/Gy$, a reasonable value for the final 
migration phase in simulations that take massive particles into account (e.g., Nice model). We chose
nine particles that were initially in the 2:5 MMR region and seven other from the 1:3 MMR region. These 
particles show the most interesting dynamical modes, especially long periods in the hibernating 
mode, like those in Figures \ref{f7} and \ref{f8}. Then, we took 123 points for each particle,
along their original simulations and restarted them by considering a migrating Neptune. Therefore, 
1107 (9x123) runs  were made for 2:5 MMR particles and 861 (7x123) runs for 1:3 MMR particles. 
In the end, $240$ particles escaped the 2:5 MMR  and $168$ particles escaped the 1:3 MMR. Figures
\ref{f11} and \ref{f12} exhibit histograms for the original eccentricities distribution 
and for the averaged eccentricities after the MMR escapes, for particles that became fossilized 
 for the 2:5 and 1:3 MMRs, respectively. It is noteworthy that particles 
that escaped the MMRs mostly
have low eccentricities. It is found that the most probable values for the 
eccentricities are 0.12 and 0.18 ($44.9\lesssim q\lesssim 48.7au$) for the 2:5 MMR case, and 
0.21 ($48.8\lesssim q\lesssim 49.3au$) for the 1:3 MMR case. These results 
confirm what is proposed by \citet{gomes2011} that orbits similar to $2004XR_{190}$ may be formed 
close to other MMRs.

 The previous experiments show that it is possible to form detached objects fossilized close to
the 2:5 and 1:3 MMRs. However, these experiments do not allow calculating the fraction of 
objects that become fossilized from the initial population, since they were chosen on purpose based 
on specific particles evolution in which the hibernating mode appeared. To make 
a more unbiased analysis that takes  the total time which the particles spend in
 the hibernating mode into account, we performed another experiment. It consists of taking the simulations with 
Neptune at $29.8 au$ and  restart them from different times, considering that Neptune migrates
 with $V_{mig}=0.5au/Gy$. All bodies present at the selected times are considered. Thus it
 is possible to verify the formation of fossilized objects in a scenario in which several types of scattered particles are perturbed by a migrating
Neptune. We consider that an object
 is fossilized if its average semimajor axis, after Neptune reaches its current position, is in
 the region $54.6<a<55.4au$ for bodies close to the 2:5 MMR or $61.8<a<62.4au$ for 
bodies close to the 1:3 MMR. A total of 15 simulations were performed. In all of them at least 
one fossilized object with $q>40au$ was formed in both regions, as shown by Table \ref{t1}. However, we note that 
these simulations are also able to produce fossilized objects with moderate perihelion distance
($35<q<40au$). This result is qualitatively in accordance with Figures \ref{f11} and \ref{f12}, which show two peaks in the distributions of detached objects, one for moderate perihelion distance and another one for high perihelion distance. Of course those figures do not show quantitative coherence since the initial conditions were chosen on purpose from orbital evolutions exhibiting the hibernating mode. Approximately 70 $\%$  of fossilized objects with moderate perihelion
distance turned out to be stable for periods of billions of years on later simulations, while 100$\%$ of
fossilized objects with high perihelion distance ($q > 40au$) are stable.
The simulations in 
this new experiment generated, on average, 4.22 fossilized particles with moderate perihelion distance
 for each  fossilized object with high perihelion distance for the 2:5 MMR case, while this ratio 
is 2.43 to 1 for the 1:3 MMR. We continued the integration of some of the particles for an extra Gy 
and found out that on average 30\% of the moderate perihelion distance 
ones did not keep stable. Thus the corrected rates not including the 
unstable particles should be 2.95/1 (2:5 MMR) and 1.70/1 (1:3 MMR).

\begin {table}
\caption{Particles fossilized with moderate and high {\it q} when Neptune achieves its current position. Left side 
stands for the results of the 2:5MMR runs, while the right side for the 1:3MMR.
The columns represent the run ID, the number of objects fossilized with $q>40au$ ($N_1$),
and the number of objects with $35<q<40au$ ($N_2$).}
    \label{t1}
\centering
\begin {tabular} {cccc||cccc}

\hline

Run & N$_1$ & N$_2$ & & Run & $N_1$ & $N_2$\\ 
\hline
$\#1$ & $5$  & $14$  &  & $\#1$ & $3$  & $5$  &  \\
$\#2$ & $2$  & $18$  &  & $\#2$ & $4$  & $8$  &  \\
$\#3$ & $5$  & $15$  &  & $\#3$ & $3$  & $8$  &  \\
$\#4$ & $2$  & $9$  &  & $\#4$ & $5$  & $11$  &  \\
$\#5$ & $1$  & $8$  &  & $\#5$ & $5$  & $7$  &  \\
$\#6$ & $4$  & $12$  &  & $\#6$ & $4$  & $10$  &  \\
$\#7$ & $4$  & $12$  &  & $\#7$ & $4$  & $10$  &  \\
$\#8$ & $3$  & $25$  &  & $\#8$ & $2$  & $11$  &  \\
$\#9$ & $6$  & $23$  &  & $\#9$ & $3$  & $10$  &  \\
$\#10$ & $6$  & $24$  &  & $\#10$ & $3$  & $12$  &  \\
$\#11$ & $6$  & $20$  &  & $\#11$ & $1$  & $8$  &  \\
$\#12$ & $8$  & $21$  &  & $\#12$ & $3$  & $10$  &  \\
$\#13$ & $5$  & $12$  &  & $\#13$ & $6$  & $3$  &  \\
$\#14$ & $4$  & $19$  &  & $\#14$ & $4$  & $8$  &  \\
$\#15$ & $6$  & $51$  &  & $\#15$ & $8$  & $20$  &  \\

            \hline

Tot.   & $67$ & $283$ &  & Tot.    & $58$ & $141$ &  \\

\end{tabular}
 \end{table}

\begin{figure}
\resizebox{\hsize}{!}{\includegraphics{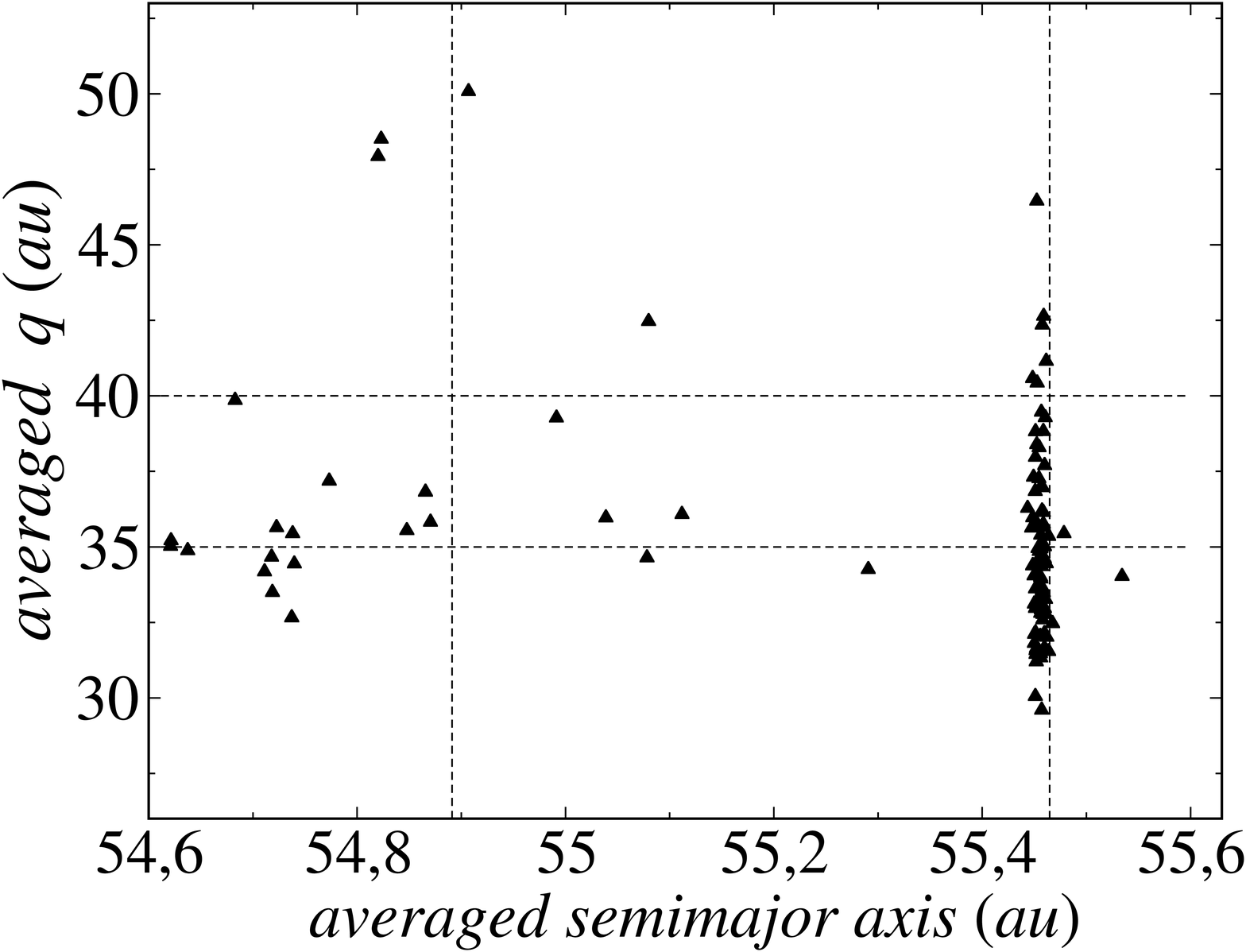}}
 \caption{ Results in the region of the 2:5 MMR for one original simulation restarted from $\sim3.5Gy$. 
Neptune was initially 
at $a_N=29.8$au and was made to migrate with $V_{mig}=0,5au/Gy$. The vertical lines account 
for the initial (at left) and final (at right) resonant semimajor axis $a_{2:5}$. All particles that 
were present in the original integration at that time they were considered.}

\label{f13}
\end{figure}

\begin{figure}
\resizebox{\hsize}{!}{\includegraphics{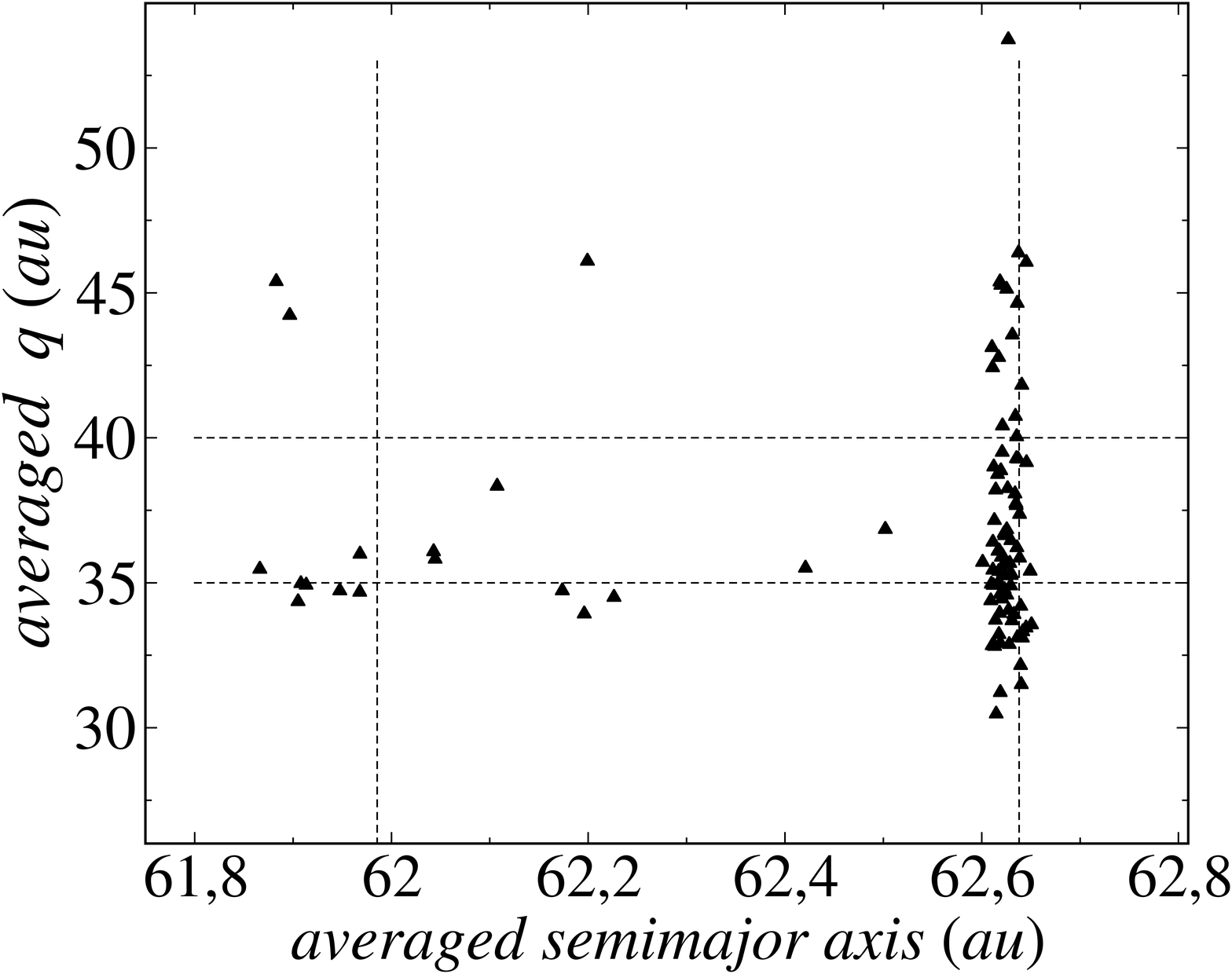}}
 \caption{ Results in the region of the 1:3 MMR for one original simulation restarted from $\sim1.0Gy$. 
Neptune was initially 
at $a_N=29.8$au and was made to migrate with $V_{mig}=0.5au/Gy$. The vertical lines account 
for the initial (at left) and final (at right) resonant semimajor axis $a_{1:3}$. All particles that 
were present in the original integration at that time were considered.}
 \label{f14}
\end{figure}

Figure \ref{f13} shows the results for the region of the 2:5 MMR for one original simulation 
without migration, which was restarted with
Neptune's migration from 
approximately 3.5 Gy, and  
Figure \ref{f14} presents the 1:3 MMR region for the same original simulation restarted from 
approximately 1Gy. Most bodies tends to remain 
trapped to the MMRs and follow the migration of Neptune. Nevertheless, some particles escape 
the resonance becoming scattered objects (below the lower horizontal line), fossilized 
with moderate perihelion distances (between horizontal lines and $ 54.6 <a <55.4au$ or $ 61.8 <a <62.4 au$)
 or  fossilized with high perihelion distance (above the upper horizontal line and $ 54.6 <a < 55.4au$ or $ 61.8 <a
 <62.4au$).

The scattered objects are very moble in the semimajor axis. They are strongly perturbed by Neptune
 and usually end up being ejected from the solar system on hyperbolic orbits. The fossilized 
objects with
 moderate perihelion distances have less mobility in semimajor axis than the scattered ones, and 
most of them are 
stable after Neptune arrives at its 
current position. The  fossilized objects with high perihelion distance, on the other hand, have almost no mobility in 
semimajor axis and are 
very stable, surviving by the age of the solar system after Neptune reaches its current 
position. Through an experiment using a more generic approach, we thus show that detached 
 objects close to the  2:5 and 1:3 MMRs with Neptune can be generated through gravitational 
interactions among 
scattered particles and giant planets without the need of an external agent (e.g., stellar 
passages or 
planetary companion with planetary mass). We just need a scattered particle to be captured in a MMR with Neptune, to experience the Kozai resonance, and to access the hibernating mode
while Neptune is still migrating to its current position.  

One comment is in order with respect to the particles that remain trapped in the MMRs with Neptune in this last experiment. Since these particles amount to many more that escape, one could try to estimate the ratio of current trapped particles to escaped ones. But we do not think this experiment is suitable for that calculation. In fact, the integration with migration lasts just 600 My in view of the initial position of Neptune and the migration speed. This is justifiable since the migration is in fact exponentially damping and we are taking it as linear, and most of the migration will take place in these first 600 My. There will therefore be some 3.5 Gy left for evolution time, during which we expect that most of the resonant particles will be scattered out of the resonance.

 Since we considered an initial uniform distribution of inclinations between $0^{\circ}$ and $50^{\circ}$, a natural question is how the formation of fossilized detached objects depends on the initial inclination. For this we considered all integrations without migration and plot histograms of the initial inclination of all particles that experienced at least one instance of a hibernating mode. Figure 
\ref{r5} shows these histograms for the 2:5 and 1:3 resonances. It is interesting to compare this with Figure \ref{r7}, which gives the distribution of inclinations near these resonances for the numerical integration presented in Section \ref{s1}. The comparison of these two distributions will be used to better estimate the mass likely deposited near the resonances in the next section. It is interesting to note that the histogram for the 1:3 resonance is much more skewed to the left than the one for the 2:5 resonance. Although we do not attempt to explain this feature, we can confirm that it is real in the sense that a particle trapped in the 1:3 MMR with a low inclination can eventually get into Kozai resonance and experience a large amplitude couple variation of the eccentricity and inclination that brings the eccentricity to near $0.1$ and the inclination to near $30^{\circ}$. Figure \ref{r6} shows an example of such behavior.

\subsection{Estimation of the mass of detached fossilized objects}

 Although the simulations  performed here were not particularly aimed at computing the total number of detached objects that would come from a past hibernating episode, we do think that we can roughly estimate the total mass that would be deposited near the 2:5 and 1:3 resonances as fossilized  objects with high perihelion distance. We consider the integrations without migration and assume that every instance of a hibernating mode creates a fossilized  object with high perihelion distance, which is reasonable in view of the examples shown in Figures\ref{f9} and \ref{f10}. For the 2:5 resonance we find that $0.38$\% of the particles show at least one instance of a hibernating mode for the first $0.85$ Gy, whereas for the 1:3 resonance this fracion is $0.55$\%. Now we compare this fraction with the total mass available in the neighborhood of each resonance taken from the Nice model simulation described in Section \ref{s1}. This mass at $0.75$ Gy, after which Neptune experiences  an effective residual migration for $0.85$ Gy is 0.041 and 0.036 Earth masses  for the 2:5 and 1:3 resonances, respectively. This yields final total masses for  fossilized objects with high perihelion near those resonances as $0.07$ and $0.09$ in 
Pluto's mass. 

If we consider the last 3$Gy$ of planetary migration, the
fractions of particles that show at least one instance of a hibernating
mode is 1.8\% and 2.2\% for the 2:5 and 1:3 MMR,
respectively. Considering the probability of escaping from
the hibernating mode during the last 3 $Gy$ is around 40\% as
suggested by Figure 15, we estimate a total mass in the neighborhood
of the 2:5 and 1:3 MMR as 0.21 and 0.26 in Pluto’s
mass.

\begin{figure}
\resizebox{\hsize}{!}{\includegraphics{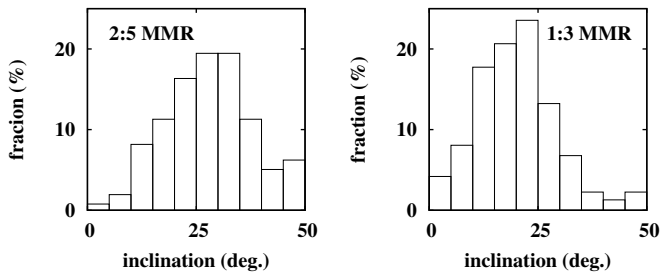}}
 \caption{Distribution of initial inclinations of particles that yielded at least one instance of a hibernating mode episode}
 
\label{r5}
\end{figure}

\begin{figure}
\resizebox{\hsize}{!}{\includegraphics{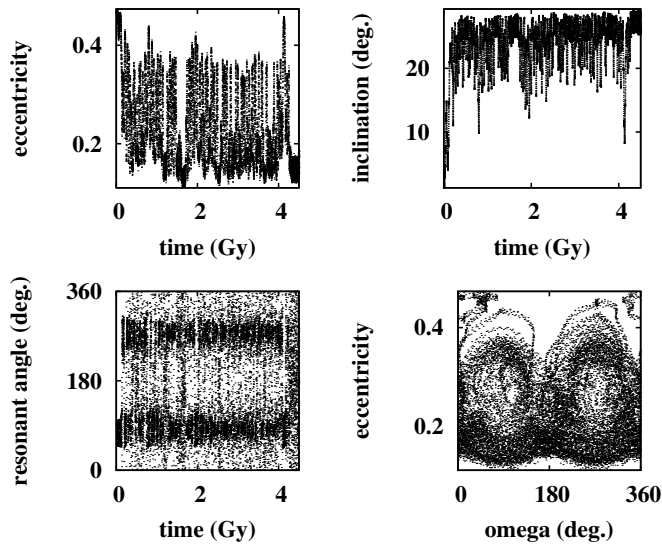}}
 \caption{Evolution of a particle trapped in the 1:3 MMR with Neptune, experiencing the Kozai resonance and eventually the hibernating mode. The particle starts with a small inclination and shows large coupled variations of the eccentricity and inclination.}
 \label{r6}
\end{figure}

Now we must also consider that this calculation was based on a uniform distribution of the inclination from $0^{\circ}$ to $50^{\circ}$. The distribution of inclinations that yields at least one instance of a hibernating mode is shown in Figure \ref{r5}. On the other hand, the expected distribution of inclinations around the 2:5 and 1:3 MMR is shown in Figure \ref{r7}. To get more accurate fractions of particles that show at least one instance of a hibernating mode, we must multiply the above numbers by a factor which is the weighed sum of the fraction of particles in each bin of the distributions shown in Figure \ref{r5}. The weights are given by normalizing the fractions given in Figure \ref{r7} so that the sum of these weights equals $10$, the number of bins. With this in mind we found the factors $0.88$ and $1.57$ that must multiply the  estimated masses above in the neighborhood of the 2:5 and 1:3 MMR's. The larger multiplying factor obtained for the 1:3 MMR reflects that the distribution of inclinations that yielded hibernating modes in Figure \ref{r5} is quite similar to the distribution of inclinations that particles should have near the 1:3 resonance according to Figure \ref{r7}.

It is also helpful to consider the great simulation presented in Section 2. As noted earlier, there are one case for the 2:5 resonance and another one for the 1:3 resonance where a particle is trapped into MMR and Kozai resonance and eventually escape those resonances. Each of these particles carries roughly $0.32$ Pluto's mass, which could be considered as another mass estimate at low eccentricities in the neighborhood of these resonances. This number is more in accordance with the mass estimate taken from the cases where we consider a longer time range during which a particle can enter and escape a hibernating mode. On the other hand, from the two examples taken from the Nice model numerical simulation, the particles that get detached near the 2:5 and 1:3 resonances are fossilized between $0.8$ and $0.9$ Gy, thus in the first $0.15$ Gy of Neptune's residual migration. We notice, however, particles that are fossilized much later coming from resonances of higher order and farther from Neptune. One caveat in the computation of the mass through the indirect method described above is that this computation is based on the total mass (particles) at the beginning of the integration. Since in the simulations undertaken in Sects. 4 and 5 we impose artificial limits for discarding the particles the number of particles in the neighborhood of the resonances decrease much faster than reality (for instance, considering the numerical integration of Section 2) which also entails a fewer particles available to experience all the mechanism that produces the fossilized  detached particles thus underestimating the total mass near those resonances. Considering all these factors, we can roughly estimate the amount of mass with low eccentricity near each of those resonances as $0.1$ to $0.3$ Pluto's mass.

\section{Conclusions}

 \label{s5}

We addressed the main aspects of the dynamical 
formation of detached objects close to the 2:5 and 1:3 MMRs with Neptune from primordial 
scattered disk particles. Our simulations show that a considerable fraction of the scattered disk particles at the neighborhood of those resonances
reach $q> 40 au$ at some point in their orbital evolution. For such an increase
 in perihelion, considering only the gravitational perturbation of the giant planets, it is 
necessary that the particle get captured in exterior mean motion resonance with Neptune and 
thereafter experience the Kozai resonance, which produces large variations in the inclination and 
perihelion. Thus it is possible that several TNOs are currently undergoing this resonance coupling, 
showing characteristics of detached objects. But sooner or later, the perihelion will decrease 
and the object may, again, be severely perturbed by Neptune.

As in \citet{gomes2011}, we also identified the emergence of the hibernating mode on particles that 
suffer the coupling between MMR + KR. This mode is characterized by long periods ($t> 100My$) 
of low eccentricity ($q> 40au$) and high inclination and can be accessed when the amplitude of
 oscillation of the resonant angle, $\phi$, becomes very high ($> 100^{\circ}$). Through the 
semi-analytical approach, for the cases of 2:5 and 1:3, we have shown the topological 
changes in the energy level curves associated with the hibernating mode. However, if there is 
no dissipative mechanism, the particle can return to experience the large amplitude anti-phase variation of the  
eccentricity and inclination characteristic of MMR+KR.

Through experiments considering the residual migration of Neptune, we show that the hibernating
 mode is a preponderant factor in the formation of fossilized objects with high 
perihelion, outside the resonant semimajor axis and without the possibility of suffering the 
Kozai mechanism again. Besides these high perihelion fossilized particles we also found  objects with 
moderate perihelion distance ($35 <q <40au$) through numerical experiments. They are not associated with the hibernating mode. We estimate that the ratio of the number of moderate-to-high perihelion objects fossilized near the 2:5 and 1:3 resonances are  2.95:1 and 1.70:1, respectively. It is important to note however that these ratios must not represent real observations since observational bias  makes it easier to observe objects with smaller perihelia. We also roughly estimated the amount of mass with low eccentricity near either of those resonances as $0.1$ to $0.3$ Pluto's mass.

As shown here for the resonances 2:5 and 1:3, and 3:8 by \citet{gomes2011}, the same 
mechanism could act in other MMRs in the trans-Neptunian region, forming other groups of 
detached objects, mostly for MMRs with $a <100au$. In principle, since there are some MMR with Neptune in the classical Kuiper belt (e.g., 4:7, 3:5, 5:8), it is possible that the same 
process could also be effective near those resonances and contribute to the formation of part of hot classical Kuiper 
belt objects (HCKBOs). We can anticipate that in preliminary tests, the hibernating mode 
showed up for some of the resonances in the classical belt. However, determining the efficiency
 of the process, as well as its relative contribution to the HCKBOs group, deserves further 
investigation that is going to be the subject of future works.

\begin{acknowledgements}
      PIOB acknowledges supports from FAPESP (grants 2011/08540-9 \& 2012/23719-8) and 
      RSG thanks CNPq for grant 301878/2007-2.
\end{acknowledgements}

\bibliographystyle{aa} 
\bibliography{references-AA-article-Brasil-et-al-ok} 



\end{document}